\documentclass{article}%
\usepackage{amsmath}
\usepackage{amsfonts}
\usepackage{amssymb}
\usepackage{graphicx}
\usepackage{epstopdf}
\usepackage[colorlinks]{hyperref}
\usepackage{amsmath,hyperref}
\usepackage{cite}
\usepackage{hyperref}
\usepackage{url}
\newcommand{\bpartial}{\mathop{\partial\kern -4pt\raisebox{.8pt}{$|$}}}

\newcommand{\bra}{\mathopen{[\kern-1.6pt[}}
\newcommand{\ket}{\mathclose{]\kern-1.5pt]}}
\newcommand{\bbra}{\mathopen{[\kern-2.2pt[\kern-2.3pt[}}
\newcommand{\bket}{\mathclose{]\kern-2.1pt]\kern-2.3pt]}}

\makeindex
\begin{document}
\title{\bf Integrable sigma model with generalized $\mathcal{F}$ structure, Yang-Baxter sigma model with generalized complex structure and multi-Yang-Baxter sigma model}
\author { A. Rezaei-Aghdam \hspace{-3mm}{ \footnote{e-mail: rezaei-a@azaruniv.ac.ir}\hspace{2mm}
{\small and }
A. Taghavi \hspace{-3mm}{ \footnote{ e-mail: amanetaghavi@gmail.com}}\hspace{2mm}} \\
{\small{\em Department of Physics, Faculty of Basic science, Azarbaijan  Shahid Madani University }}\\
{\small{\em  53714-161, Tabriz, Iran  }}}

\maketitle
\begin{abstract}
We construct an integrable sigma model with a generalized $\mathcal{F}$ structure, which involves a generalized Nijenhuis structure $\mathcal{J}$ satisfying $\mathcal{J}^{3}=-\mathcal{J}$. Utilizing the expression of the generalized complex structure on the metric Lie group manifold $G$ in terms of operator relations on its Lie algebra $\mathfrak{g}$, we formulate a Yang-Baxter sigma model with a generalized complex structure. Additionally, we present multi-Yang-Baxter sigma models featuring two and three compatible Nijenhuis structures. Examples for each of these models are provided.

\end{abstract}
\newpage

\section{\bf Introduction}
Two-dimensional integrable $\sigma$ models and their deformations have garnered considerable attention over the past 45 years \cite{KP, VG, HE}. The integrable deformation of the principal chiral model on $SU(2)$ was first presented in \cite{BP, CI, VA, NJ}. Additionally, the Yang-Baxter deformation of the chiral model was introduced by Klimcik \cite{CK, CKL, CKL1}. The Yang-Baxter sigma model relies on R-operators that satisfy the (modified) classical Yang-Baxter equation ((m)CYBE) \cite{MY, KMY, KY, EP, EP1}. The integrable sigma model on a Lie group with a complex structure was also studied in \cite{BK} as a special case of the Yang-Baxter sigma model \cite{CK}. Recently, Mohammedi proposed a deformation of $WZW$ models using two invertible linear operators \cite{NMO}.
  Utilizing the general method outlined in another work by Mohammedi \cite{NM}, we have constructed an integrable sigma model on a general manifold, particularly on a Lie group with a complex structure, in our previous work \cite{RA}. Furthermore, employing this method, we have explored an integrable sigma model on Lie groups equipped with a generalized complex structure \cite{RA}.

  Following the methodology outlined in \cite{NMO} and \cite{NM}, and building upon our previous work \cite{RA}, we aim to construct integrable sigma models on a Lie group with generalized $\mathcal{F}$ structure \cite{YA} (a generalized Nijenhuis structure $\mathcal{J}$ \cite{HI,GU} with $\mathcal{J}^{3}=-\mathcal{J}$). By utilizing the expression of the generalized complex structure on a metric Lie group $\mathbf{G}$ in terms of operator relations on its Lie algebra $\mathfrak{g}$ (as discussed in \cite{RA}), we will further develop the Yang-Baxter sigma model with generalized complex structure. Consequently, we will present the multi-Yang-Baxter sigma model. The plane of the paper is outlined as follows:

  In Section 2, we will review the methods outlined in \cite{NMO} and \cite{NM}, as well as the concepts related to generalized complex structure \cite{HI,GU}.
Then, in Section 3, we will present the integrable sigma model on a Lie group with generalized $\mathcal{F}$ structure. We will illustrate this with two examples: one on the Heisenberg Lie group $\mathbf{H_{4}}\equiv \mathbf{A_{4,8}}$ \cite{PJ}, and another on the Lie group $\mathbf{GL(2,R)}$.
In Section 4, by employing the expression of the generalized complex structure on a metric Lie group $G$ in terms of operator relations on its Lie algebra (as discussed in our previous work \cite{RA}), we will construct the Yang-Baxter sigma model with generalized complex structure. An example on the Nappi-Witten Lie group ($\mathbf{A_{4,10}}$) \cite{NW} will be presented.
Finally, in Section 5, we will introduce the multi-Yang-Baxter sigma model with two and three compatible Nijenhuis structures and provide corresponding examples.

\section{\bf Some Review on integrable sigma model and generalized complex structure}
Before delving into our main discussions, let's provide a brief overview of the two methods proposed by Mohammedi for constructing integrable sigma models \cite{NMO, NM}. Additionally, we'll briefly review the fundamental concepts and relations concerning generalized complex structure, as discussed in the literature \cite{HI, GU}.

\subsection{\bf First Method}\label{sub21}

 Consider the following sigma model action\footnote{As in \cite{NMO} and \cite{NM} instead of using the world sheet coordinates $\tau$ and $\sigma$, we will use the complex coordinates ($z=\sigma+i\tau,\bar{z}=\sigma-i\tau$) with $\partial=\frac{\partial}{\partial z}$ and $\bar{\partial}=\frac{{\partial}}{\partial\bar{z}}$. We also adopt the convention that the Levi-Civita tensor $\epsilon^{z\bar{z}}=1$; so we will not have $i$ in front of $B$ field in (\ref{eqn:sm}).}
\begin{equation}\label{eqn:sm}
S=\int_{\Sigma}\hspace{2mm}dzd\bar{z}(G_{\mu\nu}(x)+B_{\mu\nu}(x))\hspace{0mm}\partial\hspace{0mm}x^{\mu}\bar{\partial}x^{\nu},
\end{equation}
where $x^{\mu}(z,\bar{z})$ $(\mu=1,2,...,d)$ are coordinates of $d$ dimensional manifold $M$, such that $G_{\mu\nu}$ and $B_{\mu\nu}$ are invertible metric and anti-symmetric tensor fields on $M$; the $(z,\bar{z})$ are coordinates of the world-sheet $\Sigma$. The equations of motion for this model have the following form \cite{NM}
\begin{equation}\label{eom}
  \bar{\partial} \partial\hspace{0cm}x^{\lambda}+\Omega^{\lambda}\hspace{0cm}_{\mu\nu}\partial\hspace{0cm}x^{\mu}\bar{\partial}x^{\nu}=
  0 \hspace{0.1cm},
  \end{equation}
where
\begin{equation}
 \Omega^{\lambda}\hspace{0cm}_{\mu\nu}=\Gamma^{\lambda}\hspace{0cm}_{\mu\nu}-H^{\lambda}\hspace{0cm}_{\mu\nu}
\hspace{0.1cm} ,
\end{equation}
such that $\Gamma^{\lambda}\hspace{0cm}_{\mu\nu}$ is the
Christoffel coefficients and components of the torsion are given by
\begin{equation}
H^{\lambda}\hspace{0cm}_{\mu\nu}=\frac{1}{2}G^{\lambda\eta}(\partial_{\eta}B_{\mu\nu}+\partial_{\nu}B_{\eta\mu}+\partial_{\mu}B_{\nu\eta})\hspace{0.1cm}
.
\end{equation}
As in \cite{NM} one can construct a linear system whose consistency
conditions are equivalence to
the equations of motion (\ref{eom}) as follows:
\begin{equation}\label{lax1}
[\partial+\partial\hspace{0cm}x^{\mu}\alpha_{\mu}(x)]\psi=0 \hspace{0.1cm},
\end{equation}
\begin{equation}\label{lax2}
 [\bar{\partial}+\bar{\partial}x^{\mu}\beta_{\mu}(x)]\psi=0\hspace{0.1cm}
 ,
\end{equation}
where the matrices $\alpha_{\mu}$ and $\beta_{\mu}$ are functions of coordinates $x^{\mu}$.
The compatibility condition of this linear system yields the
equations of motion, provided that the matrices $\alpha_{\mu}(x)$
and $\beta_{\mu}(x)$ must satisfies the following relations \cite{NM}
\begin{equation}
 \beta_{\mu}-\alpha_{\mu}=\mu_{\mu}\hspace{0.1cm} ,
\end{equation}
\begin{equation}
 \partial\hspace{0cm}_{\mu}\beta_{\nu}-\partial\hspace{0cm}_{\nu}\alpha_{\mu}+[\alpha_{\mu},\beta_{\nu}]=\Omega^{\lambda}\hspace{0cm}_{\mu\nu}\mu_{\lambda}
\hspace{0.1cm} ,
\end{equation}
such that the second equation of the above set can then be rewritten
as
\begin{equation}\label{me}
 F_{\mu\nu}=-(\nabla_{\mu}\mu_{\nu}-\Omega^{\lambda}\hspace{0cm}_{\mu\nu}\mu_{\lambda})\hspace{0.1cm}
 ,
\end{equation}
 where the field strength $F_{\mu\nu}$ and covariant derivative corresponding to the matrices $\alpha_{\mu}$ are given as follows:
\begin{equation}
 F_{\mu\nu}=\partial_{\mu}\alpha_{\nu}-\partial_{\nu}\alpha_{\mu}+[\alpha_{\mu},\alpha_{\nu}]\hspace{1cm},\hspace{1cm}
 \nabla_{\mu}X=\partial_{\mu}X+[\alpha_{\mu},X]\hspace{0.1cm} .
\end{equation}
 Then by splitting symmetric and anti-symmetric parts of (\ref{me}), we have \cite{NM}
\begin{equation}\label{eqn:sp}
 0=\nabla_{\mu}\mu_{\nu}+\nabla_{\nu}\mu_{\mu}-2\Gamma^{\lambda}\hspace{0cm}_{\mu\nu}\mu_{\lambda}\hspace{0.1cm} ,
\end{equation}
\begin{equation}\label{eqn:asp}
 F_{\mu\nu}=-\frac{1}{2}(\nabla_{\mu}\mu_{\nu}-\nabla_{\nu}\mu_{\mu})-H^{\lambda}\hspace{0cm}_{\mu\nu}\mu_{\lambda}\hspace{0.1cm}
 .
\end{equation}
 In this manner the integrability condition of the sigma model (\ref{eqn:sm}) is equivalent to finding the matrices $\alpha_{\mu}$ and $\mu_{\mu}$ such that they satisfy
 in (\ref{me}) (or (\ref{eqn:sp}) and (\ref{eqn:asp})).

\subsection{\bf Second Method}
In \cite{NMO} Mohammedi give a general integrable deformation of chiral and WZW models.  Here we review his results as the following proposition:

\vspace{0.5cm}  $\mathbf{Proposition}$ \cite{NMO}: Consider invertible linear operators $\mathcal{P}'$ and $\mathcal{Q}'$ (such that $\mathcal{P}'+\mathcal{Q}'$ is also invertible) on a metric Lie algebra $\mathfrak{g}$ (with ad invariant invertible metric\footnote{The ad invariant metric is obtained as \cite{NW} \begin{equation*}\label{adg}
f^{\gamma}\hspace{0cm}_{\alpha\beta}g_{\gamma\delta}+f^{\gamma}\hspace{0cm}_{\alpha\delta}g_{\gamma\beta}=0\hspace{0.1cm},\end{equation*}
where the $f^{\gamma}\hspace{0cm}_{\alpha\beta}$ are the structure constants of the Lie algebra $\mathfrak{g}$ with bases $\{T_{\alpha}\}$, such that $[T_{\alpha},T_{\beta}]=f^{\gamma}\hspace{0cm}_{\alpha\beta}T_{\gamma}$.
} $< \hspace{1mm},\hspace{1mm} >_{\mathfrak{g}}$) with Lie group $G$, such that the operators satisfy the following relations \cite{NMO}:
 \begin{equation}\label{px}
\forall X,Y \in \mathfrak{g}:\hspace{1cm} < \mathcal{P'}X,Y>_{\mathfrak{g}}=<X, \mathcal{Q'}Y>_{\mathfrak{g}}-2\lambda < \mathcal{P'}X, \mathcal{Q'}Y>_{\mathfrak{g}},
\end{equation}
 \begin{equation}\label{py}
[ \mathcal{P'}X,\mathcal{Q'}Y]-\mathcal{P'}[X,\mathcal{Q'}Y]-\mathcal{Q'}[\mathcal{P'}X,Y]=\frac{\varepsilon}{2}(\mathcal{P'}+\mathcal{Q'})[X,Y],
\end{equation}
with $ \mathcal{P'}^{-1}=\mathcal{Q'}^{-t}+2\lambda I$, where $I$ is identity operator; $\lambda$ and $\varepsilon$ are parameters. Then the following two-dimensional sigma model is classically integrable \cite{NMO}:
\begin{equation*}
  S(g)=\lambda \int_{\partial B} dzd\bar{{z}}<g^{-1} \partial g,g^{-1} \bar{\partial} g >_{\mathfrak{g}}  +\frac{\lambda}{6} \int_ {B} d^3x \epsilon^{\mu\nu\rho}<g^{-1} \partial_{\mu} g,[g^{-1} \partial_{\nu},g^{-1} \partial_{\rho}]>_{\mathfrak{g}}
 \end{equation*}
  \begin{equation}\label{QM}
  +  \int_{\partial B} dzd\bar{z}<g^{-1} \partial g,{\mathcal{Q'}}^{-1}(g^{-1} \bar{\partial} g) >_{\mathfrak{g}},
  \end{equation}
where $B$ is a three-dimensional manifold with boundary ${\partial B}$. Indeed the equations of motion of the above model are given as conservation of the following currents:
 \begin{equation*}
J=g({\mathcal{P}'}^{-1}(g^{-1}\partial g))g^{-1},
 \end{equation*}
 \begin{equation}
\bar{J}=g({\mathcal{Q}'}^{-1}(g^{-1}\bar{\partial} g))g^{-1},
 \end{equation}
 i.e.
 \begin{equation}
\partial \bar{J}+\bar{\partial} J=0.
 \end{equation}
Such that these currents satisfy the zero curvature condition
\begin{equation}
\partial \bar{J}-\bar{\partial} J+\varepsilon[J,\bar{J}]=0,
 \end{equation}
if $\mathcal{P}'$ and $\mathcal{Q}'$ satisfy relation (\ref{py}) \cite{NMO}.

 \subsection{\bf Review on generalized complex structure}
 The generalized complex structure \cite{HI,GU} $\mathcal{J}$ on a manifold ${M}$ with even
dimension is an endomorphism from $\mathcal{J}:{TM}\oplus{T^*M}\rightarrow{TM}\oplus {{T}^*{M}}$, such that this structure
is invariant with respect to inner product ${\langle }\hspace{1mm} ,\hspace{1mm} {\rangle}$ on ${TM}\oplus {{T}^*{M}}$:
\begin{equation}
\forall X,Y \in {TM}\hspace{0.1cm}, \hspace{0.1cm}\xi,\eta \in {{T}^*{M}}:\hspace{1cm}\langle {\mathcal{J}}(X+\xi) , \mathcal{J}(Y+\eta)\rangle=\langle
X+\xi,Y+\eta\rangle\hspace{1mm},
\end{equation}
and
\begin{equation}\label{JJ}
\mathcal{J}^2=-1,
\end{equation}
also the generalized Nijenhuis tensor with Courant bracket \cite{CU}
\begin{equation}\label{tm}
[X+\xi , Y+\eta]_{C}=[X,Y]+L_{X}{\eta}-L_{Y}{\xi}-\dfrac{1}{2}d_M[i_{X}\eta-i_{Y}\xi],
\end{equation}
is zero  \cite{HI,GU} .i.e.
\begin{equation}\label{gnj}
N_\mathcal{J}(X+\xi,Y+\eta)=[X+\xi,Y+\eta]_{C}+\mathcal{J}[X+\xi,\mathcal{J}(Y+\eta)]_{C}+\mathcal{J}[\mathcal{J}(X+\xi),(Y+\eta)]_{C}-[\mathcal{J}(X+\xi),\mathcal{J}(Y+\eta)]_{C}=0.
\end{equation}
Utilizing the generalized complex structure $\mathcal{J}$ in the following block form \cite{CU}
\begin{equation}\label{Jb}
\mathcal{J}=\left(
\begin{array}{cccc}
  J & P \\
  Q &-J^* \\
 \end{array}\right),
 \end{equation}
where ${J}{\in}{C}^{\infty}({TM}\otimes{T^*M})$ ,
 ${P}{\in}{C}^{\infty}({\wedge^{2}{TM}})$, ${Q}\in{C}^{\infty}({\wedge}^{2}{T^*M})$; then in the coordinate representation $\{x^\mu\}$ of the manifold $M$ we have $J=J^{\mu}\hspace{0cm}_{\nu}\partial_{\mu}\otimes dx^{\nu}$,
 $P=P^{\mu\nu}\partial_{\mu}\wedge \partial_{\nu}$
 and $Q=Q_{\mu\nu}dx^{\mu}\wedge dx^{\nu}$ such that by applying the above block form (\ref{Jb}) in (\ref{JJ}), we have the following relations for tensors $J$, $P$ and $Q$:
\begin{equation}\label{pp}
P^{\nu\\k}+P^{k\nu}=0\hspace{0.5cm} ,\hspace{0.5cm}Q_{\nu\\k}+Q_{k\nu}=0,
\end{equation}
\begin{equation}
J^{\nu}\hspace{0cm}_{\mu}J^{\mu}\hspace{0cm}_{k}+P^{\nu\mu}Q_{{\mu}{k}}+\delta^{{\nu}}\hspace{0cm}_{k}=0,
\end{equation}
\begin{equation}\label{jp}
J^{\nu}\hspace{0cm}_{\mu}P^{\mu\\k}+J^{k}\hspace{0cm}_{\mu}P^{\mu\nu}=0 ,
\end{equation}
\begin{equation}\label{jq}
Q_{\nu\mu}J^{\mu}\hspace{0cm}_{k}+Q_{k\mu}J^{\mu}\hspace{0cm}_{\nu}=0.
\end{equation}
 Furthermore, relation (\ref{gnj}) (zero of the generalized Nijenhuis tensor) leads to the following relations \cite{RZ}
\begin{equation}\label{A}
\mathcal{A}^{\nu\\k\mu}=P^{\nu\lambda}\partial_{\lambda}P^{k\mu}+P^{k\lambda}\partial_{\lambda}P^{\mu\nu}+P^{\mu\lambda}\partial_{\lambda}P^{\nu\\k}=0\hspace{0.1cm}
,
\end{equation}
\begin{equation}
\mathcal{B}^{k\mu}\hspace{0cm}_{\nu}=J^{\lambda}\hspace{0cm}_{\nu}\partial_{\lambda}P^{k\mu}+P^{k\lambda}(\partial_{\nu}J^{\mu}\hspace{0cm}_{\lambda}-\partial_{\lambda}J^{\mu}\hspace{0cm}_{\nu})-P^{\mu\lambda}(\partial_{\nu}J^{k}\hspace{0cm}_{\lambda}-\partial_{\lambda}J^{k}\hspace{0cm}_{\nu})-\partial_{\nu}(J^{k}\hspace{0cm}_{\lambda}P^{\lambda\mu})=0\hspace{0.1cm}
,
\end{equation}
\begin{equation}
\mathcal{C}_{\nu\\k}\hspace{0cm}^{\mu}=J^{\lambda}\hspace{0cm}_{\nu}\partial_{\lambda}J^{\mu}\hspace{0cm}_{k}-J^{\lambda}\hspace{0cm}_{k}\partial_{\lambda}J^{\mu}\hspace{0cm}_{\nu}-J^{\mu}\hspace{0cm}_{\lambda}\partial_{\nu}J^{\lambda}\hspace{0cm}_{k}+J^{\mu}\hspace{0cm}_{\lambda}\partial_{k}J^{\lambda}\hspace{0cm}_{\nu}+P^{\mu\lambda}(\partial_{\lambda}Q_{\nu\\k}+\partial_{\nu}Q_{k\lambda}+\partial_{k}Q_{\lambda\nu})=0\hspace{0.1cm}
,
\end{equation}
\begin{equation*}
 \mathcal{D}_{\nu k
  \mu}=J^{\lambda}\hspace{0cm}_{\nu}(\partial_{\lambda}Q_{k\mu}+\partial_{k}Q_{\mu\lambda}+\partial_{\mu}Q_{\lambda
  k})+J^{\lambda}\hspace{0cm}_{k}(\partial_{\lambda}Q_{\mu\nu}+\partial_{\mu}Q_{\nu\lambda}+\partial_{\nu}Q_{\lambda\mu})+J^{\lambda}\hspace{0cm}_{\mu}(\partial_{\lambda}Q_{\nu
  k}+\partial_{\nu}Q_{k\lambda}+\partial_{k}Q_{\lambda\nu})
\notag\end{equation*}
\begin{equation}\label{D}
-\partial_{\nu}(Q_{k\lambda}J^{\lambda}\hspace{0cm}_{\mu})-\partial_{k}(Q_{\mu\lambda}J^{\lambda}\hspace{0cm}_{\nu})-\partial_{\mu}(Q_{\nu\lambda}J^{\lambda}\hspace{0cm}_{k})=0.
\end{equation}
So the components of a generalized complex structure (\ref{Jb}) on a manifold $M$ must be satisfy relations (\ref{pp}-\ref{D}).

 \section{\bf Integrable sigma model with generalized $\mathcal{F}$ structure}
 In our previous work \cite{RA}, we constructed integrable sigma models with generalized complex structures, i.e., structures that satisfy relations (\ref{pp}-\ref{D}). Here, we will consider a generalized $\mathcal{F}$ structure, i.e., instead of condition (\ref{JJ}) (or (\ref{pp}-\ref{jq})), we consider the $\mathcal{F}$ structure \cite{YA}
 \begin{equation}
 \mathcal{J}^3+\mathcal{J}=0.
 \end{equation}
So by using of block form  (\ref{Jb}) in the coordinate representation, we have the following conditions (instead of (\ref{pp}-\ref{jq})):
 \begin{equation}\label{YBe1}
P^{\mu\lambda}Q_{\lambda\rho}J^{\rho}\hspace{0cm}_{\nu}+J^{\mu}\hspace{0cm}_{\lambda}P^{\lambda\rho}Q_{\rho\nu}-P^{\mu\lambda}J^{\rho}\hspace{0cm}_{\lambda}Q_{\rho\nu}+J^{\mu}\hspace{0cm}_{\lambda}J^{\lambda}\hspace{0cm}_{\rho}J^{\rho}\hspace{0cm}_{\nu}+J^{\mu}\hspace{0cm}_{\nu}=0,
\end{equation}
\begin{equation}\label{ybe2}
J^{\mu}\hspace{0cm}_{\lambda}J^{\lambda}\hspace{0cm}_{\rho}P^{\rho\nu}+P^{\mu\lambda}Q_{\lambda\rho}P^{\rho\nu}+P^{\mu\lambda}J^{\rho}\hspace{0cm}_{\lambda}J^{\nu}\hspace{0cm}_{\rho}-J^{\mu}\hspace{0cm}_{\lambda}P^{\lambda\rho}J^{\nu}\hspace{0cm}_{\rho}+P^{\mu\nu}=0,
\end{equation}
\begin{equation}\label{YBe3}
Q_{\mu\lambda}J^{\lambda}\hspace{0cm}_{\rho}J^{\rho}\hspace{0cm}_{\nu}-J^{\lambda}\hspace{0cm}_{\mu}Q_{\lambda\rho}J^{\rho}\hspace{0cm}_{\nu}+Q_{\mu\lambda}P^{\lambda\rho}Q_{\rho\nu}+J^{\lambda}\hspace{0cm}_{\mu}J^{\rho}\hspace{0cm}_{\lambda}Q_{\rho\nu}+Q_{\mu\nu}=0.
\end{equation}
Now for the case that $M$ is a Lie group $G$ (with $ad$ invariant metric) by using the veilbein formalism  ${{\forall}{g}\in{G}},g^{-1}\partial g=(g^{-1}\partial{g})^{\alpha}T_{\alpha}
= e^{\alpha}\hspace{0cm}_{\mu}{\partial}x^{\mu}T_{\alpha}$ with inverse $e_{\beta}\hspace{0cm}^{\mu}e^{\alpha}\hspace{0cm}_{\mu}=\delta^{\alpha}\hspace{0cm}_{\beta}$ \cite{NA} and
\begin{equation}
J^{\mu}\hspace{0cm}_{\nu}=e_{\alpha}\hspace{0cm}^{\mu}J^{\alpha}\hspace{0cm}\hspace{0cm}_{\beta}e^{\beta}\hspace{0cm}_{\nu},\hspace{0.5cm}P^{\mu\nu}=e_{\alpha}\hspace{0cm}^{\mu}P^{\alpha\beta}e_{\beta}\hspace{0cm}^{\nu}, \hspace{0.5cm} Q_{\mu\nu}=e^{\alpha}\hspace{0cm}_{\mu}Q_{\alpha\beta}e^{\beta}\hspace{0cm}_{\nu},\end{equation}
 the above relations can be rewritten as follows:
\begin{equation}\label{ybe1}
P^{\alpha\delta}Q_{\delta\sigma}J^{\sigma}\hspace{0cm}_{\beta}+J^{\alpha}\hspace{0cm}_{\delta}P^{\delta\sigma}Q_{\sigma\beta}-P^{\alpha\delta}J^{\sigma}\hspace{0cm}_{\delta}Q_{\sigma\beta}+J^{\alpha}\hspace{0cm}_{\delta}J^{\delta}\hspace{0cm}_{\sigma}J^{\sigma}\hspace{0cm}_{\beta}+J^{\alpha}\hspace{0cm}_{\beta}=0\hspace{1mm},
\end{equation}
\begin{equation}\label{ybe2}
J^{\alpha}\hspace{0cm}_{\delta}J^{\delta}\hspace{0cm}_{\sigma}P^{\sigma\beta}+P^{\alpha\delta}Q_{\delta\sigma}P^{\sigma\beta}+P^{\alpha\delta}J^{\sigma}\hspace{0cm}_{\delta}J^{\beta}\hspace{0cm}_{\sigma}-J^{\alpha}\hspace{0cm}_{\delta}P^{\delta\sigma}J^{\beta}\hspace{0cm}_{\sigma}+P^{\alpha\beta}=0\hspace{1mm},
\end{equation}
\begin{equation}\label{ybe3}
Q_{\alpha\delta}J^{\delta}\hspace{0cm}_{\sigma}J^{\sigma}\hspace{0cm}_{\beta}-J^{\delta}\hspace{0cm}_{\alpha}Q_{\delta\sigma}J^{\sigma}\hspace{0cm}_{\beta}+Q_{\alpha\delta}P^{\delta\sigma}Q_{\sigma\beta}+J^{\delta}\hspace{0cm}_{\alpha}J^{\sigma}\hspace{0cm}_{\delta}Q_{\sigma\beta}+Q_{\alpha\beta}=0\hspace{1mm},
\end{equation}
Furthermore, for ease of computation, we assume that $J$, $P$, and $Q$ satisfy relations (\ref{jp}) and (\ref{jq}). Therefore, we will consider (\ref{jp}) and (\ref{jq}) in the following algebraic forms \cite{RA}
\begin{equation}\label{jpa}
J^{\alpha}\hspace{0cm}_{\delta}P^{\delta\beta}+J^{\beta}\hspace{0cm}_{\delta}P^{\delta\alpha}=0 ,
\end{equation}
\begin{equation}\label{jqa}
Q_{\alpha\delta}J^{\delta}\hspace{0cm}_{\beta}+Q_{\beta\delta}J^{\delta}\hspace{0cm}_{\alpha}=0.
\end{equation}
 In the same way the relations (\ref{A}-\ref{D}) can be rewritten as follows \cite{RA}
\begin{equation}\label{gce1}
A^{\alpha\beta\gamma}=f^{\alpha}\hspace{0cm}_{\delta\sigma}P^{\beta\sigma}P^{\gamma\delta}+f^{\gamma}\hspace{0cm}_{\delta\sigma}P^{\beta\delta}P^{\alpha\sigma}+f^{\beta}\hspace{0cm}_{\delta\sigma}P^{\alpha\delta}P^{\gamma\sigma}=0
\hspace{1mm}, \end{equation}
\begin{equation}
B^{\beta\gamma}\hspace{0cm}_{\alpha}=f^{\delta}\hspace{0cm}_{\sigma\alpha}P^{\beta\sigma}J^{\gamma}\hspace{0cm}_{\delta}+f^{\delta}\hspace{0cm}_{\alpha\sigma}P^{\gamma\sigma}J^{\beta}\hspace{0cm}_{\delta}+f^{\gamma}\hspace{0cm}_{\sigma\delta}P^{\beta\delta}J^{\sigma}\hspace{0cm}_{\alpha}+f^{\beta}\hspace{0cm}_{\sigma\delta}P^{\gamma\sigma}J^{\delta}\hspace{0cm}_{\alpha}=0\hspace{1mm}
,
\end{equation}
\begin{align}
C^{\alpha}\hspace{0cm}_{\beta\gamma}=-f^{\delta}\hspace{0cm}_{\beta\gamma}J^{\alpha}\hspace{0cm}_{\sigma}J^{\sigma}\hspace{0cm}_{\delta}-f^{\delta}\hspace{0cm}_{\beta\gamma}P^{\alpha\sigma}Q_{\sigma\delta}-f^{\alpha}\hspace{0cm}_{\delta\sigma}J^{\delta}\hspace{0cm}_{\beta}J^{\sigma}\hspace{0cm}_{\gamma}-f^{\delta}\hspace{0cm}_{\gamma\sigma}J^{\alpha}\hspace{0cm}_{\delta}J^{\sigma}\hspace{0cm}_{\beta}+f^{\delta}\hspace{0cm}_{\beta\sigma}J^{\alpha}\hspace{0cm}_{\delta}J^{\sigma}\hspace{0cm}_{\gamma}
\notag\end{align}
\begin{equation}
+f^{\delta}\hspace{0cm}_{\sigma\beta}P^{\alpha\sigma}Q_{\delta\gamma}+f^{\delta}\hspace{0cm}_{\gamma\sigma}P^{\alpha\sigma}Q_{\delta\beta}=0\hspace{1mm},
\end{equation}
\begin{equation}\label{gce4}
D_{\alpha\beta\gamma}=f^{\delta}\hspace{0cm}_{\alpha\sigma}J^{\sigma}\hspace{0cm}_{\beta}Q_{\delta\gamma}+f^{\delta}\hspace{0cm}_{\gamma\sigma}J^{\sigma}\hspace{0cm}_{\beta}Q_{\alpha\delta}+f^{\delta}\hspace{0cm}_{\gamma\sigma}J^{\sigma}\hspace{0cm}_{\alpha}Q_{\delta\beta}+f^{\delta}\hspace{0cm}_{\beta\sigma}J^{\sigma}\hspace{0cm}_{\alpha}Q_{\gamma\delta}+f^{\delta}\hspace{0cm}_{\beta\sigma}J^{\sigma}\hspace{0cm}_{\gamma}Q_{\delta\alpha}+f^{\delta}\hspace{0cm}_{\alpha\sigma}J^{\sigma}\hspace{0cm}_{\gamma}Q_{\beta\delta}=0\hspace{1mm},
\end{equation}
Note that for obtaining relations (\ref{gce1}-\ref{gce4}) we have used the Maurer-Cartan relation \cite{NA}

\begin{equation}\label{mo}
f^{\gamma}\hspace{0.0cm}_{\alpha\beta}=e^{\gamma}\hspace{0.0cm}_{\nu}(e_{\alpha}\hspace{0.0cm}^{\mu}\partial_{\mu}e_{\beta}\hspace{0.0cm}^{\nu}-e_{\beta}\hspace{0.0cm}^{\mu}\partial_{\mu}e_{\alpha}\hspace{0.0cm}^{\nu})
.\end{equation}
\subsection{The model}
Now, we will attempt to consider an integrable sigma model on the Lie group $G$ with the aforementioned generalized $\mathcal{F}$ structure, subject to conditions (\ref{ybe1}-\ref{gce4}). We propose the action of the sigma model with generalized $\mathcal{F}$ structure as follows:
\begin{align}
S=\int(g^{-1}\partial
g)^{\alpha}\{g_{\alpha\beta}+kg_{\alpha\delta}J^{\delta}_{\beta}+k^{'}Q_{\alpha\beta}+k^{''}g_{\alpha\delta}P^{\delta\sigma}g_{\sigma\beta}+k_{1}g_{\alpha\delta}J^{\delta}\hspace{0cm}_{\sigma}J^{\sigma}\hspace{0cm}_{\beta}+k_{2}g_{\alpha\delta}P^{\delta\sigma}Q_{\sigma\beta}\notag\end{align}
\begin{equation}\label{gcybm}
+k_{3}g_{\alpha\delta}J^{\delta}\hspace{0cm}_{\sigma}P^{\sigma\gamma}g_{\gamma\beta}+k_{4}Q_{\alpha\delta}J^{\delta}\hspace{0cm}_{\beta}\}(g^{-1}\partial
g)^{\beta}dzd\bar{z},
\end{equation}
where $k,k',k'',k_{1},...,k_{4}$ are real coupling constants.
To obtain the condition under which the above model is integrable, one can utilize the first method mentioned in subsection \ref{sub21} by introducing the matrices $\alpha_{\mu}$ and $\mu_{\mu}$ as follows:
\begin{equation*}
{\alpha}_{\mu}=(\lambda_{1}J^{\alpha}\hspace{0cm}_{\gamma}+\lambda_{2}P^{\alpha\delta}g_{\delta\gamma}+\lambda_{3}g^{\alpha\delta}Q_{\delta\gamma}+\lambda_{4}J^{\alpha}\hspace{0cm}_{\delta} J^{\delta}\hspace{0cm}_{\gamma}+\lambda_{5}P^{\alpha\delta}Q_{\delta\gamma}+\lambda_{6}J^{\alpha}\hspace{0cm}_{\delta}P^{\delta\sigma}g_{\sigma\gamma}+\lambda_{7}g^{\alpha\sigma}Q_{\sigma\delta}J^{\delta}\hspace{0cm}_{\gamma}  )e^{\gamma}\hspace{0cm}_{\mu}T_{\alpha}\hspace{0.1cm},
\end{equation*}
\begin{equation}\label{abgc}
{\mu}_{\mu}=(\lambda'_{1}J^{\alpha}\hspace{0cm}_{\gamma}+\lambda'_{2}P^{\alpha\delta}g_{\delta\gamma}+\lambda'_{3}g^{\alpha\delta}Q_{\delta\gamma}+\lambda'_{4}J^{\alpha}\hspace{0cm}_{\delta} J^{\delta}\hspace{0cm}_{\gamma}+\lambda'_{5}P^{\alpha\delta}Q_{\delta\gamma}+\lambda'_{6}J^{\alpha}\hspace{0cm}_{\delta}P^{\delta\sigma}g_{\sigma\gamma}+\lambda'_{7}g^{\alpha\sigma}Q_{\sigma\delta}J^{\delta}\hspace{0cm}_{\gamma}  )e^{\gamma}\hspace{0cm}_{\mu}T_{\alpha}\hspace{0.1cm},
 \end{equation}
 here $(\lambda_{1},...,\lambda_{7},\lambda'_{1},...,\lambda'_{7})$ are constant parameters and $T_{\alpha}$ are the matrix representation of the bases of Lie algebra. Since this process is quite intricate, it's preferable to perform it for examples individually. Here, we will consider the model (\ref{gcybm}) for the Heisenberg Lie group $\mathbf{H}{4}\equiv\mathbf{A_{4,8}}$ and the Lie group $\mathbf{GL(2,R)}$.
\subsection{\bf Examples}
$\mathbf{a}$) Here, we construct the integrable sigma model with generalized $\mathcal{F}$ structure (\ref{gcybm}) on the Heisenberg Lie group $\mathbf{H}{4}\equiv\mathbf{A_{4,8}}$ with the following commutation relations for its Lie algebra $A_{4,8}$ \cite{PJ}:
\begin{equation}\label{h41}
[P_{1},P_{2}]=P_{2}\hspace{0.5cm},\hspace{0.5cm}[P_{1},J]=-J\hspace{0.5cm},\hspace{0.5cm}[J,P_{2}]=T.
\end{equation}
To calculate the vierbein we parameterize the corresponding Lie group $\mathbf{H_{4}}$ with coordinates $x^{\mu}=\{x,y,u,v\}$ and using the $g$ elements as follows
\begin{equation}\label{h42}
g=e^{vT_{4}}e^{uT_{3}}e^{xT_{1}}e^{yT_{2}},
\end{equation}
where we use the generators $T_{\alpha} = \{P_{1}, P_{2}, J, T\}$. Thus the veirbein and ad-invariant metric
 have the following form \cite{EP,ER}
\begin{equation}\label{h43}
(e^{\alpha}\hspace{0.0cm}_{\mu})=\left(
\begin{array}{cccc}
  1 & 0&0&0 \\
  y &1&0&0 \\
  0&0&e^x&0\\
  0&0&y e^x&1\\
 \end{array}\right)\hspace{0.1cm},\hspace{0.1cm}(g_{\alpha\beta})=\left(
\begin{array}{cccc}
 m&0&0&-k_{0} \\
  0 &0&k_{0}&0 \\
  0&k_{0}&0&0\\
 -k_{0}&0&0&0\\
 \end{array}\right),
\end{equation}
where $k_{0}\in \mathbb{R}-\{0\}$ is an arbitrary real constant. Now, by utilizing the conditions for the generalized $\mathcal{F}$ structure (\ref{ybe1}- \ref{ybe3}) and (\ref{jpa}-\ref{jqa}), along with its integrability conditions (\ref{gce1}-\ref{gce4}), one can obtain the following results:
\begin{align}
J^{\alpha}\hspace{0.0cm}_{\beta}=\left(
\begin{array}{cccccc}
  a &0&0&0 \\
 0&0&0&0\\
 0&b&a&0\\
  -\frac{ad}{c}&0&0&0\\
 \end{array}\right)\hspace{0.1cm},\hspace{0.1cm} P^{\alpha\beta}=\left(
\begin{array}{cccccc}
  0 &0&c&0\\
 0&0&0&0 \\
  -c&0&0&d\\
0&0&-d&0\\
 \end{array}\right)
 \notag\end{align}
 \begin{equation}
Q_{\alpha\beta}=\frac{(1+a^2)}{c}\left(
\begin{array}{cccccc}
 0 &\frac{b}{a} &1&0 \\
 -\frac{b}{a} &0&0&0 \\
 -1&0&0&0\\
 0&0&0&0\\
 \end{array}\right),
\end{equation}
where $b,d \in \mathbb{R}$ and $a,c\in \mathbb{R}-\{0\}$. Here, for simplicity, we will choose $b=d=0$ and $a=c=1$. Then, the action (\ref{gcybm}) of this model, after omitting surface terms, can be written as:\footnote{ Note that, here and in the following examples, we have omitted the surface terms.}
\begin{equation*}
S=\int\hspace{2mm}dz d\bar{z}\{m(1+r)\partial x\bar{\partial} x-\frac{k_{0}}{2}(2+r)(\partial v\bar{\partial} x+\partial x\bar{\partial}v)+\frac{k_{0}}{2}(2+r)e^x(\partial y\bar{\partial} u+\partial u\bar{\partial}y)
\notag\end{equation*}
 \begin{equation*}
-k_{0}e^x(ry+k_{0}sy^2)(\partial u\bar{\partial} x-\partial x\bar{\partial}u)-\frac{k_{0}}{2}(r+2k_{0}sy)e^x(\partial u\bar{\partial}y-\partial y\bar{\partial}u)
\notag\end{equation*}
  \begin{equation} \label{yba48}
-k_{0}^{2}sy(\partial v\bar{\partial} x-\partial x\bar{\partial}v)\},
\end{equation}
where $r=k+k_{1}-2k_{2}$ and $s=k''+k_{3}$. The model (\ref{yba48}) satisfies the integrability condition (\ref{me}) if we set
  \begin{equation*}
\lambda_{1}=-\frac{1}{s}(\lambda_{6}+\lambda_{2}-s(1-\lambda_{4}+2\lambda_{5})),\lambda_{3}=-\lambda_{7},\lambda'_{3}=-\lambda'_{7},
 \end{equation*}
  \begin{equation*}
\lambda'_{1}=\frac{(\lambda_{2}+\lambda_{6})(2+r)(\lambda_{2}+\lambda_{6}-(1+\lambda'_{4}-2\lambda'_{5})s)+(\lambda'_{4}-2\lambda'_{5})s^2}{((\lambda_{2}+\lambda_{6})(2+r)-s)s},
  \end{equation*}
   \begin{equation}\label{a481con}
\lambda'_{2}=-\frac{(\lambda_{2}+\lambda_{6})(\lambda_{2}+\lambda_{6}+\lambda'_{6})(2+r)-\lambda'_{6}s}{(\lambda_{2}+\lambda_{6})(2+r)-s},
 m=0, \end{equation}

with $\lambda_{2},\lambda_{4},\lambda_{5},\lambda_{6},\lambda_{7},\lambda'_{4},\lambda'_{5},\lambda'_{6},\lambda'_{7}$ are arbitrary constants and one of them can be considered as spectral parameters. It should be explained that
 by applying the parameter $m$ from (\ref{a481con}) in the model (\ref{yba48}), we obtain the following components of the Christoffel symbols and torsion \footnote{Note that, in examples $\mathbf{a}$ and $
\mathbf{b}$, the Christoffel and torsion components of the models (\ref{yba48}) and (\ref{8rm}), which we apply in the integrability condition (\ref{me}) are very lengthy. Therefore, we will express them after finding the integrability conditions.}
  \begin{equation}\label{cha48}
\Gamma^{2}\hspace{0cm}_{12}=\frac{1}{2},\hspace{0.5cm}\Gamma^{3}\hspace{0cm}_{13}=\frac{1}{2},\hspace{0.5cm}\Gamma^{4}\hspace{0cm}_{23}=\frac{e^x}{2},
 \end{equation}
  \begin{equation*}
  H^{1}\hspace{0cm}_{12}=\frac{k_{0}s}{2+r},\hspace{0.5cm}H^{2}\hspace{0cm}_{12}=-\frac{2k_{0}sy+r}{2(2+r)},\hspace{0.5cm}H^{3}\hspace{0cm}_{13}=\frac{2k_{0}sy+r}{2(2+r)},
 \end{equation*}
  \begin{equation}\label{ta48}
  H^{3}\hspace{0cm}_{14}=\frac{k_{0}se^{-x}}{2+r},\hspace{0.5cm}H^{4}\hspace{0cm}_{23}=\frac{(2k_{0}sy+r)e^{x}}{2(2+r)},\hspace{0.5cm}H^{4}\hspace{0cm}_{24}=\frac{k_{0}s}{2+r}.
 \end{equation}
To check the integrability of the model (\ref{yba48}), one can consider that equation of motion (\ref{eom}) for sigma model (\ref{yba48}) with Christoffel and torsion components (\ref{cha48}) and (\ref{ta48}), can be written as linear equations (\ref{lax1}) and (\ref{lax2}). On the other hand, the $WZW$ (Witten-Zumino-Witten) model on a Lie group $\mathbf{G}$ is defined as follows:
 \begin{equation}\label{wzwt}
S_{WZW}=\frac{K}{4\pi}\int_{\Sigma}dz d\overline{z}\hspace{1mm} e^{\alpha}\hspace{0cm}_{\mu}g_{\alpha \beta}e^{\beta}\hspace{0cm}_{\nu}\hspace{1mm} \partial x^{\mu}\overline{\partial}x^{\nu}+\frac{K}{24\pi}\int_{B}\hspace{2mm}d^3
\sigma\varepsilon^{i j k}e^{\alpha}\hspace{0cm}_{\mu}g_{\alpha \delta}f^{\delta}\hspace{0cm}_{\beta \gamma}e^{\beta}\hspace{0cm}_{\nu}e^{\gamma}\hspace{0cm}_{\lambda}\partial_{i}x^{\mu}\partial_{j}x^{\nu}\partial_{k}x^{\lambda},
\end{equation}
so the $WZW$ model on Heisenberg Lie group $\mathbf{H_{4}}$ (with $K = 4\pi$) is given by the following action
 \cite{EP,ER}:
\begin{equation}\label{a48wzw}
S_{WZW_{A48}}=\int {dz}d\bar{z}\{m\partial x\bar{\partial}x-k_{0}(\partial x\bar{\partial}v +\partial v\bar{\partial}x)+k_{0} e^{x}(\partial y\bar{\partial}u+\partial u\bar{\partial}y+y\partial u\bar{\partial}x-y\partial x\bar{\partial}u)\},
\end{equation}
by setting $m=0$ from (\ref{a481con}); the action (\ref{yba48}) can be considered as the following integrable deformation of the $WZW$ action
 \begin{equation*}
 S=S_{WZW_{A48}}+k_{0}\int {dz}d\bar{z}\{-\frac{r}{2}(\partial v\bar{\partial} x+\partial x\bar{\partial}v)+\frac{r}{2}e^x(\partial y\bar{\partial} u+\partial u\bar{\partial}y)
 \end{equation*}
\begin{equation}
-(1+\frac{r}{2}) ye^x(\partial u\bar{\partial}x-\partial x\bar{\partial}u)-\frac{k_{0}}{2}sy^2e^x(\partial u\bar{\partial} x-\partial x\bar{\partial}u)-k_{0}sy(\partial v\bar{\partial} x-\partial x\bar{\partial}v)\}.
\end{equation}
Note that, as we expected, this deformation is different from Yang-Baxter deformation of the $WZW$ model of Heisenberg Lie group $H_{4}$ table 1 of ref \cite{EP}. Indeed, the third, fourth, and fifth terms in the second part of the action (59) are new terms.
\vspace{0.5cm}

$\mathbf{b}$) As second example, we construct integrable sigma model with generalised $\mathcal{F}$ structure on the $\mathbf{GL(2,R})$ Lie group. By considering the following commutation relations between the generators $\{T_{1},T_{2},T_{3},T_{4}\}$  \cite{EP1,NW}
\begin{equation}\label{h41}
[T_{1},T_{2}]=2T_{2}\hspace{0.5cm},\hspace{0.5cm}[T_{1},T_{3}]=-2T_{3}\hspace{0.5cm},\hspace{0.5cm}[T_{2},T_{3}]=T_{1},\hspace{0.5cm}[T_{4},...]=0\hspace{0.1cm},
\end{equation}
where $T_{4}$ is a central generator, and we parameterize the corresponding Lie group $\mathbf{GL(2,R})$ with coordinates $x^{\mu}=\{x,y,u,v\}$ using the $g$ elements as
\begin{equation}\label{h42}
g=e^{yT_{2}}e^{xT_{1}}e^{uT_{3}}e^{vT_{4}},
\end{equation}
then the veirbein and ad-invariant metric are given by \cite{EP1}
\begin{equation}\label{h43}
(e^{\alpha}\hspace{0.0cm}_{\mu})=\left(
\begin{array}{cccc}
  1 & ue^{-2x}&0&0 \\
  0 &e^{-2x}&0&0 \\
  -2u&-u^2e^{-2x}&1&0\\
  0&0&0&1\\
 \end{array}\right)\hspace{0.1cm},\hspace{0.1cm}(g_{\alpha\beta})=\left(
\begin{array}{cccc}
  2k_{0}&0&0&0 \\
  0 &0&k_{0}&0 \\
  0&k_{0}&0&0\\
 0&0&0&m\\
 \end{array}\right),
\end{equation}
where $k_{0}$, $m \in \mathbb{R}-\{0\}$. By using the conditions for the integrability of the generalized complex structure
(\ref{gce1} -\ref{gce4}) and the relations (\ref{ybe1}-\ref{jqa}),
then the following components of generalized $\mathcal{F}$ structure are obtained
\begin{align}
J^{\alpha}\hspace{0.0cm}_{\beta}=\left(
\begin{array}{cccccc}
  a &b&-\frac{bf^2}{4h^2}&0 \\
  \frac{fe}{4h^3} &\frac{f(4bh+cf)}{4h^2}& \frac{f^2e}{16h^4}&0 \\
  -\frac{cf}{h}&c&\frac{e+4bhf}{4h^2}&0\\
  -\frac{d(e-2ah^2)}{2h^3}&-\frac{d(2bh+cf)}{2h^2}& -\frac{df(2bfh+e)}{8h^4}&0\\
 \end{array}\right)\hspace{0.1cm},\hspace{0.1cm}
 P^{\alpha\beta}=\left(
\begin{array}{cccccc}
 0&\frac{f^2}{4h}&h&-\frac{df}{2h} \\
 -\frac{f^2}{4h}&0&f&-\frac{f^2d}{4h^2}\\
 -h&-f&0&d\\
  \frac{df}{2h}&\frac{f^2d}{4h^2}&-d&0\\
 \end{array}\right),
 \notag\end{align}
\begin{equation}
 Q_{\alpha\beta}=\frac{l}{h^2(bf+ah)}\left(
\begin{array}{cccccc}
  0 &c&\frac{e}{4h^2}&0 \\
  -c&0&b&0 \\
  -\frac{e}{4h^2}&-b&0&0\\
  0&0&0&0\\
 \end{array}\right),
\end{equation}
where $l=bf(2ah+bf)+h^2(1+a^2),e=-cf^2+4ah^2$ and $h,(bf+ah) \in \mathbb{R}-\{0\}$, $a,b,c,d,f \in \mathbb{R}$. Here for simplicity we will choose $b=0,a=c=d=h=1,f=2$; then one can construct sigma model (\ref{gcybm}), as follows:
\begin{equation*}
S=\int\hspace{2mm}dz d\bar{z}\{2k_{0}(1+r)\partial x\bar{\partial} x -e^{-2x}k_{0}r(1-u)(\partial x\bar{\partial}y+\partial y\bar{\partial} x)+k_{0}r(1-u)^2e^{-4x}\partial y\bar{\partial} y
\end{equation*}
\begin{equation*}
+\frac{mr}{2}(\partial x\bar{\partial}v+\partial v\bar{\partial} x)+\frac{k_{0}}{2}e^{-2x}(2+r)(\partial u\bar{\partial} y+\partial y\bar{\partial}u)-\frac{mr}{2}(1-u)e^{-2x}(\partial y\bar{\partial} v+\partial v\bar{\partial}y)+m\partial v\bar{\partial}v
\notag\end{equation*}
\begin{equation*}
+e^{-2x}(2k_{0}^2su^2-k_{0}(r+4k_{0}s)u)(\partial x\bar{\partial}y-\partial y\bar{\partial} x)+2k_{0}msu(\partial x\bar{\partial} v-\partial v\bar{\partial}x)
\notag\end{equation*}
\begin{equation} \label{8rm}
-\frac{m}{2}e^{-2x}(r+2k_{0}s-(r+4k_{0}s)u+2k_{0}su^2)(\partial v\bar{\partial} y-\partial y\bar{\partial}v)+\frac{k_{0}}{2}e^{-2x}(4k_{0}s(1-u)+r)(\partial u\bar{\partial} y-\partial y\bar{\partial}u)\}.
\end{equation}
Here, similar to the previous example, we set $r=k+k_{1}-2k_{2}$ and $s=k''+k_{3}$.
To investigate the integrability of this model such that it satisfies the relation (\ref{me}), we assume that $\alpha_{\mu}$ and $\mu_{\mu}$ are similar to the relation (\ref{abgc}). Then, by setting:
\begin{equation}\label{gl2r1con}
r=0,\hspace{0.5cm}m=-2k_{0}
\end{equation}
\begin{equation}\label{gl2r2con}
\lambda_{1}=-\lambda_{4}+2\lambda_{5},\lambda'_{1}=-\lambda'_{4}+2\lambda'_{5},\lambda_{2}=-\lambda_{6}+s,\lambda'_{2}=-\lambda'_{6}-2s,\lambda_{3}=-\lambda_{7},\lambda'_{3}=-\lambda'_{7},
\end{equation}
such that $\lambda_{4},\lambda_{5},\lambda_{6},\lambda_{7},\lambda'_{4},\lambda'_{5},\lambda'_{6},\lambda'_{7}$ are arbitrary constants (and one of them can be considered as spectral parameter), this model satisfies the integrability condition (\ref{me}). Indeed by applying  the condition (\ref{gl2r1con}) the Christoffel and torsion components of the model (\ref{8rm}) can be expressed as follows
   \begin{equation}
\Gamma^{1}\hspace{0cm}_{23}=\frac{e^{-2x}}{2},\hspace{0.5cm}\Gamma^{2}\hspace{0cm}_{12}=-1,\hspace{0.5cm}\Gamma^{3}\hspace{0cm}_{13}=-1,
 \end{equation}
  \begin{equation*}
H^{1}\hspace{0cm}_{24}=k_{0}s(1-u)^2e^{-2x},\hspace{0.2cm}H^{1}\hspace{0cm}_{34}=k_{0}s,\hspace{0.2cm}H^{2}\hspace{0cm}_{14}=-2k_{0}se^{2x},\hspace{0.2cm}H^{2}\hspace{0cm}_{24}=2k_{0}s(1-u),\hspace{0.2cm}H^{3}\hspace{0cm}_{41}=2k_{0}s(1-u)^2,
\end{equation*}
 \begin{equation}
H^{3}\hspace{0cm}_{43}=2k_{0}s(1-u),\hspace{0.2cm}H^{4}\hspace{0cm}_{21}=k_{0}s(1-u)^2e^{-2x},\hspace{0.2cm}H^{4}\hspace{0cm}_{13}=-k_{0}s,\hspace{0.2cm} H^{4}\hspace{0cm}_{32}=-k_{0}s(1-u)e^{-2x},
 \end{equation}
so the conditions (\ref{gl2r1con}) and (\ref{gl2r2con}) satisfy the integrability condition (\ref{me}) of the model (\ref{8rm}). Consider the WZW model of the $\mathbf{GL(2,R)}$ Lie group \cite{EP1}
\begin{equation}\label{a48wzw}
S_{WZW_{GL(2,R)}}=\int {dz}d\bar{z}\{m\partial v\bar{\partial}v+2k_{0}\partial x\bar{\partial}x+k_{0}e^{-2x}(\partial u\bar{\partial}y+\partial y\bar{\partial}u) +k_{0}e^{-2x}(\partial u\bar{\partial}y-\partial y\bar{\partial}u)\},
\end{equation}
if we set $r=0,m=-2k_{0}$ (from (\ref{gl2r1con})), then the action (\ref{8rm}) can be written as the perturbed action of the $WZW$ model as follows
 \begin{equation*}
 S=S_{WZW_{GL(2,R)}}+k_{0}\int {dz}d\bar{z}\{-(1+2k_{0}s)e^{-2x}(\partial u\bar{\partial}y-\partial y\bar{\partial}u)+2k_{0}su^2e^{-2x}(\partial x\bar{\partial}y-\partial y\bar{\partial} x)
\end{equation*}\begin{equation}\label{8rwzw}
-4k_{0}su(\partial x\bar{\partial} v-\partial v\bar{\partial}x)+2k_{0}se^{-2x}(1-u)^2(\partial v\bar{\partial} y-\partial y\bar{\partial}v)+2k_{0}s(1-u)e^{-2x}(\partial u\bar{\partial} y-\partial y\bar{\partial}u) \}.
\end{equation}
In comparison with the above result, the Yang-Baxter deformation of $WZW$ on $\mathbf{GL(2,R)}$ (table one of \cite{EP1}), we see that our model (\ref{8rwzw}) corresponds to the sum of two cases $iii$ and $v$ from that table. In other words, the sum of two deformations is also a deformation.
\section{\bf Yang-Baxter sigma model with generalized complex structure}
As mentioned in our previous work \cite{RA}, the generalized complex structure relations on a metric Lie group (\ref{pp}-\ref{jq}) and (\ref{gce1}-\ref{gce4}) can be expressed as operator relations by considering the operators $J$, $\mathcal{P}$, $\mathcal{Q}:\mathfrak{g} \rightarrow \mathfrak{g}$ that act on the generators $T_{\alpha}$ of the Lie algebra $\mathfrak{g}$ as follows:
 \begin{equation}\label{gco0}
{J}(T_{\alpha})={J}^{\beta}\hspace{0.0cm}_{\alpha}T_{\beta}\hspace{0.5cm},\hspace{0.5cm}\mathcal{P}(T_{\alpha})=\mathcal{P}_{\alpha}\hspace{0.0cm}^{\beta}T_{\beta}\hspace{0.5cm},\hspace{0.5cm}\mathcal{Q}(T_{\alpha})=\mathcal{Q}_{\alpha}\hspace{0.0cm}^{\beta}T_{\beta},
\end{equation}
such that we define
\begin{equation}
\mathcal{P}_{\alpha}\hspace{0.0cm}^{\beta}=g_{\alpha\gamma}P^{\gamma\beta}\hspace{0.1cm},\hspace{0.1cm}\mathcal{Q}_{\alpha}\hspace{0.0cm}^{\beta}=Q_{\alpha\gamma}g^{\gamma\beta}
,\end{equation}
where $g_{\alpha\beta}=<T_{\alpha},T_{\beta}>$ is an ad-invariant invertible metric on Lie algebra $\mathfrak{g}$. Then conditions (\ref{pp}-\ref{D}) (on a metric Lie algebra $\mathfrak{g}$) can be rewritten as the following operator relations on $\mathfrak{g}$ \cite{RA} :
\begin{equation}\label{gco1}
\mathcal{P}^{t}=-\mathcal{P}\hspace{1cm},\hspace{1cm}\mathcal{Q}^{t}=-\mathcal{Q},
\end{equation}
\begin{equation}\label{gco2}
{J}^{2}+\mathcal{P}\mathcal{Q}+I=0,
\end{equation}
\begin{equation}
({J\mathcal{P}})^{t}=-({J\mathcal{P}}),
\end{equation}
\begin{equation}
({J\mathcal{Q}})^{t}=-({J\mathcal{Q}}).
\end{equation}
Furthermore relations (\ref{gce1}-\ref{gce4}) can be rewritten on the following operator relations \cite{RA}:

$\forall X,Y \in \mathfrak{g}:$
\begin{equation}\label{gco6}
[\mathcal{P}(X),\mathcal{P}(Y)]-\mathcal{P}[X,\mathcal{P}(Y)]-\mathcal{P}[\mathcal{P}(X),Y]=0,
\end{equation}
\begin{equation}
[\mathcal{P}(X),J(Y)]+[J(X),\mathcal{P}(Y)]=-J^t([\mathcal{P}(X),Y]+[X,\mathcal{P}(Y)]),
\end{equation}
\begin{equation}
[X,Y]-[J(X),J(Y)]+J[J(X),Y]+J[X,J(Y)]+\mathcal{P}[X,\mathcal{Q}(Y)]+\mathcal{P}[\mathcal{Q}(X),Y]=0,
\end{equation}
\begin{equation}\label{gco7}
[J(X),\mathcal{Q}(Y)]+[\mathcal{Q}(X),J(Y)]+J^t([\mathcal{Q}(X),Y]+[X,\mathcal{Q}(Y)])-\mathcal{Q}([J(X),Y]+[X,J(Y)])=0,
\end{equation}
where transpose of an operator $O$ is defined as
\begin{equation}
 \forall X,Y \in \mathfrak{g}\hspace{0.1cm},\hspace{0.5cm}<OX,Y>=<X,O^tY>.
\end{equation}
 In the following, for constructing the action (\ref{QM}), we assume that $\lambda=0$, and $\mathcal{P}'$ and $\mathcal{Q}'$ are given with the following relations:
 \footnote{Note that in the previous work \cite{RA}, we have considered the model (\ref{QM}) with $\mathcal{Q}'^{-1}=kI+\xi_{1}J^t-\xi_{2}\mathcal{Q}-\xi_{3}\mathcal{P}$ and $\lambda=0$.}
  \begin{equation*}
  \mathcal{Q}'=kI+\xi_{1}J^t-\xi_{2}\mathcal{Q}-\xi_{3}\mathcal{P},
  \end{equation*}
  \begin{equation}\label{pq}
  \mathcal{P}'=kI+\xi_{1}J+\xi_{2}\mathcal{Q}+\xi_{3}\mathcal{P},
  \end{equation}
  where $\mathcal{P}'= \mathcal{Q}'^t$ . Then by inserting (\ref{pq}) in (\ref{py}) and using (\ref{gco1}-\ref{gco7}) we must have
  \begin{equation*}
  \xi_{2}^2(\mathcal{Q}[X,\mathcal{Q}(Y)]+\mathcal{Q}[\mathcal{Q}(X),Y]-[\mathcal{Q}(X),\mathcal{Q}(Y)])
  \end{equation*}
   \begin{equation*}
  +\xi_{2}\xi_{3}(\mathcal{Q}[X,\mathcal{P}(Y)]+\mathcal{Q}[\mathcal{P}(X),Y]-[\mathcal{Q}(X),\mathcal{P}(Y)]-[\mathcal{P}(X),\mathcal{Q}(Y)])
  \end{equation*}
   \begin{equation}\label{QMC}
  -\xi_{1}\xi_{3}(\mathcal{P}[J(X),Y]+\mathcal{P}[X,J(Y)])=0.
  \end{equation}
  Here, we assume that $k\varepsilon-k^2=\xi_{1}^2=\xi_{2}\xi_{3}$ and $J=-J^t$. Indeed, from (\ref{py}), the relation (\ref{QMC}) represents the integrability condition for the following Yang-Baxter sigma model \cite{CKL} with a generalized complex structure\footnote{Note that by means of Yang-Baxter sigma model with generalized complex structure is that the action (\ref{ybpqm}) construct from components of generalized complex structure $\mathcal{J}$ (\ref{Jb}) on $\mathfrak{g}\oplus \mathfrak{g}^*$ that satisfy in zero Nijenhuis (\ref{gnj}) with Courant bracket, i.e relations (\ref{gce1}-\ref{gce4}); which is equivalent to relations (\ref{gco6}-\ref{gco7}). Indeed relations (\ref{gco6}-\ref{gco7})  different from zero of Nijenhuis for $\mathcal{R}=\xi_{1}J+\xi_{2}\mathcal{Q}+\xi_{3}\mathcal{P}$ on $\mathfrak{g}$.} (with three deformation parameters)
 \begin{equation}\label{ybpqm}
 S=\int dz d\bar{{z}} <g^{-1}\partial g,\frac{1}{kI-\xi_{1}J-\xi_{2}\mathcal{Q}-\xi_{3}\mathcal{P}},g^{-1}\bar{\partial} g>.
   \end{equation}
 In the following we investigate an example for the model (\ref{ybpqm}).
 \subsection{\bf Example}

$\mathbf{c})$ As an example we construct the sigma model (\ref{ybpqm}) on Nappi-Witten Lie group $\mathbf{A_{4,10}}$, with the following Lie algebra commutators \cite{PJ,NW}
\begin{equation}
 [J,P_{1}]=P_{2}\hspace{0.5cm} ,\hspace{0.5cm}[J,P_{2}]=-P_{1}\hspace{0.5cm} ,\hspace{0.5cm}
 [P_{1},P_{2}]=T.
 \end{equation}
 We parameterize the corresponding Lie group $\mathbf{A_{4,10}}$ with coordinates $x^{\mu}=\{ a_{1} ,a_{2}, u,v\}$ and consider the following parametrization for the group element \cite{PJ}
\begin{equation}
g=e^{\Sigma a_{i}P_{i}}e^{uJ+vT}\hspace{0cm},
 \end{equation}
with the generators $T_{\alpha}=\{ P_{1}, P_{2}, J,T\}$. The veirbein and ad-invariant metric is given by \cite{NW}:
\begin{equation}\label{ga410}
(e^{\alpha}\hspace{0.0cm}_{\mu})=\left(
\begin{array}{cccc}
  cos(u)&sin(u)&0&0 \\
 -sin(u)&cos(u)&0&0\\
  0&0&1&0\\
  \frac{1}{2}a_{2}&-\frac{1}{2}a_{1}&0&1 \\
 \end{array}\right)\hspace{0.1cm},\hspace{0.1cm}(g_{\alpha\beta})=\left(
\begin{array}{cccc}
  1 &0&0&0 \\
  0 &1&0&0 \\
  0&0&k_{0}&1\\
 0&0&1&0\\
 \end{array}\right),
 \end{equation}
where $k_{0}\in \mathbb{R}$. The corresponding components of the generalized complex structure (which satisfy the algebraic form of the relations (\ref{pp}-\ref{jq}), (\ref{gce1}-\ref{gce4}) and  $J=-J^t$ ) are
\begin{equation}
(J^{\alpha}\hspace{0.0cm}_{\beta})=\left(
\begin{array}{cccc}
  0 &-1&0&0 \\
  1 &0&0&0 \\
  0&0&0&0\\
 0&0&0&0\\
 \end{array}\right)\hspace{0.1cm},\hspace{0.1cm}(Q_{\alpha\beta})=\left(
\begin{array}{cccc}
  0 &0&0&0 \\
  0 &0&0&0 \\
  0&0&0&a\\
 0&0&-a&0\\
 \end{array}\right),\hspace{0.1cm}(P^{\alpha\beta})=\left(
\begin{array}{cccc}
  0 &0&0&0 \\
  0 &0&0&0 \\
  0&0&0&\frac{1}{a}\\
0&0&-\frac{1}{a}&0\\
 \end{array}\right),
\end{equation}
 so the action (\ref{ybpqm}) for this model is given by
\begin{equation*}
 S(g)= \int dzd\tilde{z}\{\frac{2ka^2+ak_{0}^{2}(\xi_{3}-a^2\xi_{2
 })}{4(k^2a^2-(\xi_{2}a^2-\xi_{3})^2)}(a_{2}(\partial u\bar{\partial}a_{1}+\partial a_{1}\bar{\partial}u)-a_{1}(\partial a_{2}\bar{\partial}u+\partial u\bar{\partial}a_{2})+2(\partial u\bar{\partial}v+\partial v\bar{\partial}u))\end{equation*}
\begin{equation*}
+\frac{ak_{0}(\xi_{3}-a^2\xi_{2
 })}{4(k^2a^2-(\xi_{2}a^2-\xi_{3})^2)}(-a_{1}a_{2}(\partial a_{1}\bar{\partial}a_{2}+\partial a_{2}\bar{\partial}a_{1})+2a_{2}(\partial a_{1}\bar{\partial}v+\partial v\bar{\partial}a_{1})-2a_{1}(\partial a_{2}\bar{\partial}v+\partial v\bar{\partial}a_{2})
\end{equation*}
\begin{equation*}
+( a_{1}^2\partial a_{2}\bar{\partial}a_{2}
+ a_{2}^2\partial a_{1}\bar{\partial}a_{1})+4\partial v\bar{\partial}v)+\frac{ak_{0}}{ka-\xi_{2}a^2+\xi_{3}}\partial u\bar{\partial}u+\frac{k}{k^2+\xi_{1}^2}(\partial a_{1}\bar{\partial}a_{1}
+ \partial a_{2}\bar{\partial}a_{2})
\end{equation*}
\begin{equation}\label{ybgca410}
+\frac{a(k_{0}^2+2)(\xi_{3}-a^2\xi_{2
 })}{4(k^2a^2-(\xi_{2}a^2-\xi_{3})^2)}(a_{2}(\partial u\bar{\partial}a_{1}-\partial a_{1}\bar{\partial}u)+a_{1}(\partial a_{2}\bar{\partial}u-\partial u\bar{\partial}a_{2}))\},
\end{equation}
 this model satisfies the integrability condition (\ref{QMC}). Comparing the above model with the results in table 4 of ref \cite{EP}, we observe that the fourth through eighth terms are new.

\section{Multi-Yang-Baxter sigma models}
The model (\ref{ybpqm}) and its integrability conditions (\ref{QMC}) are motivated to consider Yang-Baxter sigma model with two and three (or multiple) compatible Nijenhuis structures\footnote{Note that in bi-Yang-Baxter sigma model where presented in \cite{CKL}, with Nijenhuis structures $J$ and $J_{g}=Ad_{g^{-1}}J Ad_{g}$ with $g\in G$, these structures ($J$ and $J_{g}$) automatically have zero concommitants. However, for the construction of the model (91) with three complex structure $J_{1},J_{2}$ and $J_{3}$, their concommitants must be zero. Thus, the two-Yang-Baxter sigma model with different $J_{1}$ and $J_{2}$ and zero concommitants is different from bi-Yang-Baxter sigma model with $J$ and $J_{g}$.}. If we consider $\mathcal{Q}'$ and $\mathcal{P}'$ in the following forms i.e. a two, three or in general a multi-parametric deformation of principal chiral model
\begin{equation*}
  \mathcal{Q}'=kI+ \xi_{1}J_{1}+\xi_{2}J_{2}+\xi_{3}J_{3},
  \end{equation*}
  \begin{equation}\label{jj}
  \mathcal{P}'=kI+\xi_{1}J_{1}^{t}+\xi_{2}J_{2}^{t}+\xi_{3}J_{3}^{t},
  \end{equation}
then from (\ref{QM}) (with $\lambda=0$) for the following multi-Yang-Baxter sigma model
 \begin{equation}\label{ybjjjm}
 S=\int dz d\bar{{z}} <g^{-1}\partial g,\frac{1}{kI+\xi_{1}J_{1}+\xi_{2}J_{2}+\xi_{3}J_{3}},g^{-1}\bar{\partial} g>,
   \end{equation}
the condition (\ref{px}) (by assuming $J_{i}^{t}=-J_{i}$ $i=1,2,3$) can be written as
 \begin{equation}
\forall X,Y \in \mathfrak{g}:\hspace{1cm} < (kI-\xi_{1}J_{1}-\xi_{2}J_{2}-\xi_{3}J_{3})X,Y>_{\mathfrak{g}}=<X,(kI+\xi_{1}J_{1}+\xi_{2}J_{2}+\xi_{3}J_{3})Y>_{\mathfrak{g}},
\end{equation}
and integrability condition (\ref{py}) result as the following relations
  \begin{equation}\label{ybjj1}
\omega_{i}[X,Y]-[J_{i}(X),J_{i}(Y)]+J_{i}[J_{i}(X),Y]+J_{i}[X,J_{i}(Y)]=0,\hspace{0.5cm}\omega_{i}=0,\pm1,\hspace{0.5cm}i=1,2,3,
\end{equation}
  \begin{equation*}
(\omega_{i}+\omega_{j})[X,Y]-[J_{i}(X),J_{j}(Y)]-[J_{j}(X),J_{i}(Y)]+J_{i}[J_{j}(X),Y]+J_{j}[J_{i}(X),Y]+
\end{equation*}
 \begin{equation}\label{ybjj2}
J_{i}[X,J_{j}(Y)]+J_{j}[X,J_{i}(Y)]=0,\hspace{0.5cm}i,j=1,2,3,\hspace{0.5cm}i\neq j,
\end{equation}
by assuming that
 \begin{equation}\label{jjj}
k(k+\varepsilon)+\omega_{1}\xi_{1}^2+\omega_{2}\xi_{2}^2+\omega_{3}\xi_{3}^2+(\omega_{1}+\omega_{2})\xi_{1}\xi_{2}+(\omega_{1}+\omega_{3})\xi_{1}\xi_{3}+(\omega_{2}+\omega_{3})\xi_{2}\xi_{3}=0,
\end{equation}
i.e. one can construct integrable multi-Yan-Baxter sigma model (\ref{ybjjjm}) with three compatible Nijenhuis structures (\ref{ybjj2}) that satisfy in mCYBE (\ref{ybjj1}).
One can also write the multi-Yang-Baxter sigma model (\ref{ybjjjm})  with two $J_{i}$ such that the conditions (\ref{ybjj1}) and (\ref{ybjj2}) are satisfied with $i,j=1,2$ and the conditions (\ref{jjj}) is replaced by
 \begin{equation}
k(k+\varepsilon)+\omega_{1}\xi_{1}^2+\omega_{2}\xi_{2}^2+(\omega_{1}+\omega_{2})\xi_{1}\xi_{2}=0.
\end{equation}
 Indeed the relations (\ref{ybjj1})-(\ref{ybjj2}) demonstrate that the two or three  Nijenhuis structures are compatible with zero concommitant \cite{HG}. In this manner one can construct a multi-Yang-Baxter sigma model as multi-parametric  Poisson-Lie deformation of principal sigma model with dual Lie algebra $\tilde{{\mathfrak{g}}}=\mathfrak{g}_{\xi_{1}J_{1}+\xi_{2}J_{2}+\xi_{3}J_{3}}$ \cite{CKL}; such that $\tilde{{\mathfrak{g}}}$ is a Lie algebra with Lie bracket, $\forall X,Y \in \mathfrak{g},\hspace{0.1cm}[X,Y]_{\xi_{1}J_{1}+\xi_{2}J_{2}+\xi_{3}J_{3}}=\xi_{1}[X,Y]_{J_{1}}+\xi_{2}[X,Y]_{J_{2}}+\xi_{3}[X,Y]_{J_{3}}$ where $[X,Y]_{J_{i}}=[J_{i}(X),Y]+[X,J_{i}(Y)]$. Relations (\ref{ybjj1}-\ref{jjj}) can also obtained from satisfying the combination of Nijenhuis structures $\xi_{1}J_{1}+\xi_{2}J_{2}+\xi_{3}J_{3}$ in Yang-Baxter equation.
Note that if we have $J_{i}^{2}=-1$ (i.e. complex structure $J_{i}$) and have Hermitian structure then we have $N=2$ structure and two of $J_{i}$ are independent \cite{HG}. In the following we investigate two examples on $\mathbf{IX+R}$ and $\mathbf{G_{6,23}}$ Lie groups.
\subsection{Examples}
$\mathbf{d})$ As first example we consider the compatible Nijenhuis structures $J_{1},J_{2}$ over $\mathbf{IX+R}$ Lie group \cite{AR} with the following commutation relations for their Lie algebra
 \begin{equation}
 [T_{2},T_{3}]=T_{1},\hspace{0.5cm} [T_{1},T_{3}]=-T_{2},\hspace{0.5cm} [T_{1},T_{2}]=T_{3},
 \end{equation}
 we parameterize the corresponding Lie group $\mathbf{IX+R}$ with coordinates $x_{
µ} =\{x_{1}, x_{2}, x_{3}, x_{4}\}$ and consider the
 the group element $g=e^{x_{1}T_{1}}e^{x_{2}T_{2}}e^{x_{3}T_{3}}e^{x_{4}T_{4}}$. Then the veirbein and ad-invariant metric are given by:
\begin{equation}\label{ga410}
(e^{\alpha}\hspace{0.0cm}_{\mu})=\left(
\begin{array}{cccc}
  cosx_{3}cosx_{2}&sinx_{3}&0&0 \\
  -sinx_{3}cosx_{2}&cosx_{3}&0&0 \\
  sinx_{2}&0&1&0\\
 0&0&0&1\\
 \end{array}\right)\hspace{0.1cm},\hspace{0.1cm}(g_{\alpha\beta})=\left(
\begin{array}{cccc}
  k_{0}&0&0&0 \\
  0 &k_{0}&0&0 \\
  0&0&k_{0}&0\\
0&0&0&k_{0}\\
 \end{array}\right),
 \end{equation}
where $k_{0}\in \mathbb{R}-\{0\}$.
 The compatible Nijenhuis structures $J_{1},J_{2}$
that satisfies the conditions (\ref{ybjj1}) and (\ref{ybjj2}) with $\omega_{1}=\omega_{2}=1
$ are given as follows
 \begin{equation}
(J_{1}^{\alpha}\hspace{0.0cm}_{\beta})=\left(
\begin{array}{cccc}
 0& 0&0&a \\
  0 &0&1&0 \\
  0&-1&0&0\\
  -a&0&0&0\\
 \end{array}\right)\hspace{0.1cm},\hspace{0.1cm}(J_{2}^{\alpha}\hspace{0.0cm}_{\beta})=\left(
\begin{array}{cccc}
  0 & 0&0&b \\
  0 &0&1&0 \\
  0&-1&0&0\\
  -b&0&0&0\\
 \end{array}\right),\end{equation}
where $a\neq b$; then the action (\ref{ybjjjm}) can be obtained as
 \begin{equation*}
 S(g)= \int dzd\bar{z}\{\frac{kk_{0}}{k^2+(\xi_{1}+\xi_{2})^2}((1-cos^2x_{2}cos^2x_{3})\partial x_{1}\bar{\partial}x_{1}+cos^2x_{3}\partial x_{2}\bar{\partial}x_{2}+\partial x_{3}\bar{\partial}x_{3}+\partial x_{4}\bar{\partial}x_{4} \end{equation*}
\begin{equation*}
-sinx_{3}cosx_{2}cosx_{3}(\partial x_{1}\bar{\partial}x_{2}+\partial x_{2}\bar{\partial}x_{1})+sinx_{2}(\partial x_{1}\bar{\partial}x_{3}+\partial x_{3
}\bar{\partial}x_{1}))
\end{equation*}
\begin{equation*}
+\frac{kk_{0}}{k^2+(a\xi_{1}+b\xi_{2})^2}(cos^2x_{2}cos^2x_{3}\partial x_{1}\bar{\partial}x_{1}+sin^2x_{3}\partial x_{2}\bar{\partial}x_{2}+\partial x_{4}\bar{\partial}x_{4}+sinx_{3}cosx_{2}cosx_{3}(\partial x_{1}\bar{\partial}x_{2}+\partial x_{2}\bar{\partial}x_{1}))
\end{equation*}
\begin{equation*}
-\frac{k_{0}(\xi_{1}+\xi_{2})}{k^2+(\xi_{1}+\xi_{2})^2}(sinx_{3}cosx_{2}(\partial x_{3}\bar{\partial}x_{1}-\partial x_{1
}\bar{\partial}x_{3})+sinx_{2}cosx_{3}(\partial x_{2}\bar{\partial}x_{1}-\partial x_{1
}\bar{\partial}x_{2}))
\end{equation*}
 \begin{equation}
 +\frac{k_{0}(a\xi_{1}+b\xi_{2})}{k^2+(a\xi_{1}+b\xi_{2})^2}(cosx_{2}cosx_{3}(\partial x_{4}\bar{\partial}x_{1}-\partial x_{1
}\bar{\partial}x_{4})+sinx_{3}(\partial x_{4}\bar{\partial}x_{2}-\partial x_{2
}\bar{\partial}x_{4}))\}.
 \end{equation}
\vspace{0.5cm}

$\mathbf{e})$  As another example, we construct (\ref{ybjjjm}) model on six-dimensional Lie group $\mathbf{G_{6,23}}$. The $g_{6,23}$ Lie algebra have the following commutations \cite{MO,RS}:
\begin{equation}
[T_{2},T_{3}]=T_{1}\hspace{0.5cm},\hspace{0.5cm}[T_{2},T_{6}]=T_{3}\hspace{0.5cm},\hspace{0.5cm}[T_{3},T_{6}]=T_{4}.
\end{equation}
Now using $l=e^{x_{1}T_{1}}e^{x_{2}T_{2}}e^{x_{3}T_{3}}e^{x_{4}T_{4}}e^{x_{5}T_{5}}e^{x_{6}T_{6}}$ as a Lie group element with coordinates $x^{\mu}=\{x_{1},x_{2},x_{3},x_{4},x_{5},x_{6}\}$ and the generators $\{T_{1}, T_{2}, T_{3}, T_{4}, T_{5}, T_{6}\}$, the algebraic metric and veilbeins related to this Lie algebra are given as \cite{RS}:
\begin{equation}\label{ec6}
(e^{\alpha}\hspace{0.0cm}_{\mu})=\left(
\begin{array}{cccccc}
  1 & x_{3}&0&0&0&0 \\
  0 &1&0&0&0&0 \\
  0&x_{6}&1&0&0&0\\
  0&\frac{1}{2}x_{6}^{2}&x_{6}&1&0&0\\
  0 &0&0&0&1&0 \\
  0 &0&0&0&0&1 \\
 \end{array}\right)\hspace{0.1cm},\hspace{0.1cm}(g_{\alpha\beta})=\left(
\begin{array}{cccccc}
  0 &0&0&0&0&m_{2} \\
  0 &m_{1}&0&m_{2}&m_{3}&m_{4} \\
  0&0&-m_{2}&0&0&0\\
  0&m_{2}&0&0&0&0\\
  0 &m_{3}&0&0&m_{5}&m_{6} \\
  m_{2} &m_{4}&0&0&m_{6}&m_{7} \\
\end{array}\right),
\end{equation}
where $m_{1},..., m_{7}$ are arbitrary real parameters. By using equations (\ref{ybjj1}) and (\ref{ybjj2}) with $\omega_{1}=\omega_{2}=\omega_{3}=1$ three compatible complex structure  $J_{1},J_{2},J_{3}$ obtain as follows:
\begin{equation*}
(J_{1}^{\alpha}\hspace{0.0cm}_{\beta})=\left(
\begin{array}{cccccc}
  0 & 0&0&-1&0&0 \\
  0 &0&0&0&0&1 \\
  0&0&0&0&a&0\\
  1&0&0&0&0&0\\
  0 &0&-a&0&0&0 \\
  0 &-1&0&0&0&0 \\
 \end{array}\right),\hspace{0.5cm}
 (J_{2}^{\alpha}\hspace{0.0cm}_{\beta})=\left(
\begin{array}{cccccc}
  0 & 0&0&-1&0&0 \\
  0 &0&0&0&0&1 \\
  0&0&0&0&b&0\\
  1&0&0&0&0&0\\
  0 &0&-b&0&0&0 \\
  0 &-1&0&0&0&0 \\
 \end{array}\right),\end{equation*}
 \begin{equation}
 (J_{3}^{\alpha}\hspace{0.0cm}_{\beta})=\left(
\begin{array}{cccccc}
  0 & 0&0&-1&0&0 \\
  0 &0&0&0&0&1 \\
  0&0&0&0&c&0\\
  1&0&0&0&0&0\\
  0 &0&-c&0&0&0 \\
  0 &-1&0&0&0&0 \\
 \end{array}\right),
\end{equation}
where $a\neq b\neq c$  are arbitrary real parameters. For simplicity we assume that $m_{1}=m_{3}=m_{4}=m_{6}=m_{7}=0$, thus our multi-Yang-Baxter integrable sigma model (\ref{ybjjjm}) can be expressed as:
\begin{equation*}
 S(g)= \int dzd\bar{z}\{\frac{km_{2}}{k^2+(\xi_{1}+\xi_{2}+\xi_{3})^2}((\partial x_{1}\bar{\partial}x_{6}+\partial x_{6}\bar{\partial}x_{1})+x_{3}(\partial x_{2}\bar{\partial}x_{6}+\partial x_{6}\bar{\partial}x_{2})+x^2_{6}\partial x_{2}\bar{\partial}x_{2}\end{equation*}
\begin{equation*}
+x_{6}(\partial x_{3}\bar{\partial}x_{2}+\partial x_{2}\bar{\partial}x_{3})+(\partial x_{4}\bar{\partial}x_{2}+\partial x_{2}\bar{\partial}x_{2}))-\frac{1}{k^2+(a\xi_{1}+b\xi_{2}+c\xi_{3})^2}(km_{2}x_{6}\partial x_{2}\bar{\partial}x_{2}
\end{equation*}
\begin{equation*}
+km_{2}(\partial x_{3}\bar{\partial}x_{2}+\partial x_{2}\bar{\partial}x_{3})+x_{6}(a\xi_{1}+b\xi_{2}+c\xi_{3})(m_{2}+m_{5})(\partial x_{5}\bar{\partial}x_{2}+\partial x_{2}\bar{\partial}x_{5})
\end{equation*}
\begin{equation}
+(a\xi_{1}+b\xi_{2}+c\xi_{3})(m_{2}+m_{5})(\partial x_{5}\bar{\partial}x_{3}+\partial x_{3}\bar{\partial}x_{5}))+\frac{(a\xi_{1}+b\xi_{2}+c\xi_{3})(m_{2}-m_{5})}{k^2+(a\xi_{1}+b\xi_{2}+c\xi_{3})^2}x_{6}(\partial x_{2}\bar{\partial}x_{5}-\partial x_{5}\bar{\partial}x_{2})\}.
 \end{equation}

\section{\bf Conclusion}
Following the methodology outlined in the previous work \cite{RA}, we construct a new integrable sigma model with generalized $\mathcal{F}$ structure. Additionally, utilizing the expression of the generalized complex structure on the metric Lie group in terms of operator relations on its Lie algebra, we construct the Yang-Baxter sigma model with generalized complex structure. Furthermore, we present the multi-Yang-Baxter sigma model with two or three compatible Nijenhuis structures. Investigating the Poisson-Lie symmetry of these models remains an open problem, as does their conformality up to one loop.

\subsection*{Acknowledgements}

This work has been supported by the research vice chancellor of Azarbaijan Shahid Madani University under research fund No. 1402/231.

\end{document}